\documentclass[twoside,a4paper,11pt]{article}

\usepackage{amsfonts}                   
\usepackage[psamsfonts]{amssymb}
\usepackage[bottom]{footmisc}
\usepackage{amsmath}
\usepackage{multicol}
\usepackage{a4wide}                     
\usepackage{fancyhdr}

\newcounter{eg}                         \newtheorem{eg}{Example}[section]        
\def\beg{\begin{eg}\rm}                 \def\eeg{\hfill\sq\end{eg}}                
\newcommand{\initiate}{\setcounter{equation}{0}} % To number equations
 \allowdisplaybreaks                    

\def\b#1{{\mathbb #1}}

\def\c#1{{\cal #1}}

\def\Dirac{{\raise0.09em\hbox{/}}\kern-0.69em D}

\def\ep{i\epsilon}

\def\exterior{{{\raise0.2em\hbox{$\scriptstyle\bigwedge$}}{}}}

\def\kb{i\kbar}

\def\kbar{{\mathchar'26\mkern-9muk}}

\def\lesssim{\mathrel{\hbox{\rlap{hbox{\lower8pt\hbox{$\sim$}}}\hbox{$<$}}}}

\def\sq{\hbox{\rlap{$\sqcap$}$\sqcup$}}

\def\p{\partial}                        
\def\pprime{{\prime\prime}}                
\def\tfrac #1#2{\textstyle{\frac{#1}{#2}}}
\def\dfrac #1#2{\displaystyle{\frac{#1}{#2}}}

\def\k{\kern-.1em\mathbin{,}\kern-.1em}
 
\def\hk{\kern.12em\raise-1em\hbox{$\hat{\raise1em\hbox{,}}$}\kern.12em}

\begin{document}

\title{On noncommutative spherically symmetric spaces}
\author{Maja Buri\'c $^{1}$\thanks{majab@ipb.ac.rs} \ 
                   and
        John Madore $^{2}$\thanks{madore@th.u-psud.fr} 
                   \\[15pt]$\strut^{1}$
        University of Belgrade,  Faculty of Physics, P.O. Box 44
                   \\
        SR-11001 Belgrade 
                   \\[5pt]$\strut^{2}$
        Laboratoire de Physique Th\'eorique 
                   \\
        F-91405 Orsay   
       }
 
\date{}
% 
% \fancypagestyle{firststyle}
% {
%    \fancyhf{}
%    \fancyfoot[C]{Page \thepage\ of \pageref{LastPage}}
% }
% 
% \fancypagestyle{plain}
% {%
%     \fancyhead{}
%     \fancyhead[ol]{\slshape \leftmark }
%     \fancyhead[RE]{\textit{ \nouppercase{\leftmark}} }
%     \fancyhead[LO]{\textit{ \nouppercase{\rightmark}} }
% %     \fancyfoot[ol]{Rev: \svnrev \ (\svnfilerev)}
% %     \fancyfoot[or]{\svnyear -\svnmonth -\svnday \ \svnhour:\svnminute} %
% %     \fancyfoot[el]{Rev: \svnrev \ (\svnfilerev)} \fancyfoot[er]{\svnyear -\svnmonth -\svnday \ \svnhour:\svnminute} %
%     \fancyfoot[C]{Page \thepage\  of \pageref{LastPage}}
% }

\maketitle
\vfill
\parindent 0pt

\begin{abstract}
  Two families of noncommutative extensions are given of a general
  space-time metric with spherical symmetry, both based on the matrix
  truncation of the functions on the sphere of symmetry. The first
  family uses the truncation to foliate space as an infinite set of
  spheres, is of dimension four and necessarily time-dependent; the
  second can be time-dependent or static, is of dimension five and uses
  the truncation to foliate the internal space.
\end{abstract}

\vfill
\setlength{\parskip}{15pt plus5pt minus0pt}
\setlength{\textheight}{24.0cm}

\pagestyle{plain}

\initiate
\section{Introduction}                                  \label{Iam}

Finding an operator description of realistic gravitational
configurations is an important task of any theory aiming to describe
quantum gravity. Such description, here referred to as
`quantization', would incorporate at least effectively properties of
spacetime beyond the classical regime; it could as well be
fundamental. The basic aspect of quantization is 
representation through operators: concrete representation yields 
the eigenvalues and the
eigenstates of the operators of position, metric and curvature.
Therefore, possible outcomes of quantization can be  discretization
of spacetime, if the spectra of  coordinates are discrete, or 
resolution of singularities, if the classically singular
values do not belong to the spectrum of curvature invariants.
Another important question, of interest in noncommutative geometry, is
whether the algebra of operators $\c{A}$ together with the 
calculus which supports the given gravitational
configuration can be in some sense regarded as a `noncommutative
space', that is, whether the notions as connection and  metric 
can be introduced geometrically or are just quantities defined `externally',
as fields with given properties.

In spite of the progress which has been made in the recent years in
noncommutative geometry and in quantum gravity in general, some
important results are still lacking. For physics the most interesting
spacetimes like the Schwarzschild black hole or the
Friedmann-Robertson-Walker (FRW) cosmology have not been `quantized'
yet, at least not to a common consent. Relevant or `realistic' models
have been obtained only in two dimensions and reduce essentially to
the quantum line and the fuzzy sphere.  A generalization of the
quantum line has been proposed in every dimension \cite{FadResTak89},
and in every even dimension there is a simple smooth
noncommutative space with planar symmetry defined by the constant
commutators $J^{\mu\nu}_0$  between  coordinates,
\begin{equation}
 [x^\mu,x^\nu] =\kb J^{\mu\nu} = \kb J^{\mu\nu}_0 .          \label{Moy}
\end{equation}
Usually $ J^{\mu\nu}_0 $ is taken to be nonsingular; constant $\kbar$
defines the length scale at which the effects of noncommutativity become
significant. However, these spaces are not isotropic except in 
two spatial dimensions while the important
classical configurations  have spherical symmetry. 

Therefore, if we aim to find `realistic' quantum spacetimes we need to
construct four-dimensional spaces with exact or slightly deformed
spherical symmetry, and this is the main objective of our paper. The
approach we are using is that of noncommutative differential geometry
because we believe that one should be able to describe quantum or
quantized gravity as the classical: through geometry. We shall see
that, although noncommutative spaces which we obtain here 
do not have exactly the desired classical limits, the formalism which
we use seems to be an appropriate tool.  We hope that it will be 
possible to find solutions yet closer to the classical ones by 
dimensional extension of the  algebras here analyzed.

In ordinary differential geometry a metric is introduced to measure
the distance between points. Since noncommutative geometry is
essentially without points the concept might seem in this case rather
pointless. It is possible however to carry over one of the definitions
of a metric which is used in commutative geometry and to give it a
meaning in certain noncommutative cases as a measure of distance.  In
ordinary differential geometry if $\xi$ and $\eta$ are 1-forms then
the value of the inner product $\, g(\xi \otimes \eta)$ at a given point
does not depend on the values of $\xi$ and $\eta$ at any other
point. This condition can be expressed as the bilinearity condition,
that for any function $f$
\begin{equation}
g(f\xi \otimes \eta) = f g(\xi \otimes \eta)
= g(\xi \otimes \eta f) = g(\xi \otimes \eta) f  .
\end{equation}
We assume that this condition remains valid in the noncommutative
case.  Without the bilinearity condition it is not possible to
distinguish for example in ordinary spacetime a metric which assigns a
function to a vector field in such a way that the value at a given
point depend only on the vector at that point from one which is some
sort of convolution over the entire manifold.

\initiate
\section{General formalism}                             \label{ga}

A noncommutative geometry as we define it~\cite{Mad00c}  consists of an
associative algebra $\,\c{A}$ (here referred to as `noncommutative
space') over which sits a differential calculus 
$\,\Omega^*(\c{A})$. Within this structure we single out a set of
$n$ 1-forms $\,\theta^\alpha$, a noncommutative extension of the
moving frame or $n$-bein of classical differential
geometry. Conversely one can define the differential calculus to be
such that any chosen special set of 1-forms commute with the elements
of the algebra
\begin{equation}
 [x^\mu,\theta^\alpha] =0.                              \label{a}
\end{equation}
We refer to it as a frame since it seems to be the most natural
generalization of the moving frame of Cartan. In fact, the bimodule
structure of the 1-forms uniquely determines a differential calculus
over the algebra,
\begin{equation}
 df = e_\alpha f \, \theta^\alpha.                        \label{d}
\end{equation}
 With the frame it is relatively easy at least to first order to calculate the
objects of interest in differential geometry: beyond first order more
imagination will certainly be necessary.

\subsection{Kinematics}

If instead of the 1-form $\theta^\alpha$ we had an element $f$ of the
algebra, then condition~(\ref{a}) could  be written  to first order 
as
\begin{equation}
 [x^\mu,f] = \kb J^{\mu\nu} \p_\nu f = 0  ,                            % \label{a}
\end{equation}
% This can also be written using the derivations $e_\alpha$ dual to the
% frame as
% \begin{align}
% J^{\alpha\beta} e_\beta f = 0
% \end{align}
and  states that the center of the algebra is trivial.
With the frame element however we can only state that the condition is
first order in the derivatives. One can think of it then as a constraint.
We shall find a second condition
\begin{align}
e_\alpha C^\alpha{}_{\beta\gamma} = 0                           \label{b}
\end{align}
in terms of the Ricci rotation coefficients which one can think of as
a gauge condition.  

The derivations $e_\beta$ dual to $\theta^\alpha$,
$ \, \theta^\alpha(e_\beta) =\delta^\alpha_\beta$,  are usually
assumed to be inner,
\begin{equation}
 e_\alpha f =[p_\alpha,f]  , \qquad p_\alpha \in \c{A} .  \label{cAA}
\end{equation}
This is the simplest and in some important cases~\cite{Mad00c,Buric:2006di} 
like matrix spaces the only
choice to define vector fields. Sometimes 
however the condition (\ref{cAA}) cannot be imposed: in the
case of quantized phase space of
quantum mechanics for example,  momenta $\,
 p_\alpha = -i\hbar \,  \delta^\mu_\alpha\,  \p_\mu \, $
do not belong to the algebra of coordinates.
When coordinates commute the dimension of the phase space is
necessarily twice the dimension of the configuration space. In
noncommutative geometry it need not be so: the phase space can be
identical to the configuration space, that is, `half' of its classical
analogon.  The answer to the question: how many operators generate
$\c{A}$? defines as we shall see the initial conditions of the 
problem which we are trying to solve.

The momenta necessarily satisfy a quadratic relation of the form
\begin{equation}
2P^{\alpha\beta}{}_{\gamma\delta} p_{\alpha} p_\beta 
- F^\beta{}_{\gamma\delta} p_\beta - K_{\beta\gamma} = 0.               \label{qr}
\end{equation}
This `structure equation' follows from the conditions
which we impose on the differential; it  gives the same information
about geometry of the space as the set of derivations $e_\alpha$ or 
the set of frame elements $\theta^\alpha$. The  differential 
$df$ of a function  $\, f\in\c{A}$ is given by  
\begin{equation}
df =  [p_\alpha,f]\theta^\alpha =
[p_\alpha\theta^\alpha, f]                                                                           \label{df}
\end{equation}
when the module of 1-forms is free and the frame is used as
basis. We can write to first order
\begin{equation}
P^{\alpha\beta}{}_{\gamma\delta} = \frac  12
\delta^{\alpha\beta}_{\gamma\delta}
+\ep Q^{\alpha\beta}{}_{\gamma\delta}       .    \label{pq}
\end{equation}
The $\ep$ here is the product of $\kb$ and the square~$\mu^2$ of a
macroscopic mass scale parameter $\mu$, which
we can relate for example with the Schwarzschild mass $m$ or 
with the cosmological constant $\Lambda$. As before $\kbar$ 
defines the scale of  noncommutativity through 
the commutation relations
\begin{equation}
[x^\mu, x^\nu] = \kb J^{\mu\nu}(x) ,               \label{commutator}
\end{equation}
and if we take for example that 
$\,\kbar \sim l_{Pl}^2$ and  $\,\mu^2\sim \Lambda$ we obtain 
$\,\epsilon \sim 10^{-122}$. The coefficients
$\, Q^{\alpha\beta}{}_{\gamma\delta}$, $\, F^\beta{}_{\gamma\delta}$,
$\, K_{\beta\gamma}$ are antisymmetric in the lower pair of indices while
$\, Q^{\alpha\beta}{}_{\gamma\delta}$ is symmetric in the upper
pair; they are hermitian elements of the center of the
algebra  $\c{A}$.  Formula~(\ref{qr})  is equivalent to writing  
the Ricci rotation coefficients $\, C^\gamma{}_{\alpha\beta} $,
\begin{equation}
 [e_\alpha, e_\beta] = C^\gamma{}_{\alpha\beta}\, e_\gamma          \label{RicciRot}
\end{equation}
as linear expressions in the momenta,
\begin{equation}
 C^\gamma{}_{\alpha\beta}=  F^\gamma{}_{\alpha\beta} 
- 4\ep p_\delta Q^{\gamma\delta}{}_{\alpha\beta}    .                \label{Cs}
\end{equation}
It can also be considered as the definition of a Poisson structure, to
which it is known~\cite{Mad97,CahGutHorRaw01,ArnHui11} one can
associate a curvature.
%  From expression (\ref{Cs}) for the rotation
% coefficients it follows that they must satisfy the integrability
% condition $\, e_\gamma C^\gamma{}_{\alpha \beta} =0$.  This condition
% would normally in commutative geometry be considered as a gauge
% condition.

All derivations here are taken in the semi-classical approximation
that is to leading order in $\epsilon$. This is in fact the only
possibility to do concrete calculations when one is solving a problem
like ours: to find an {\it a priori} unknown algebra, that is, 
commutators $J^{\mu\nu}(x)$. Then  we have for example
\begin{equation}
[x^\mu, f] 
= \kb J^{\mu\nu} \p_\nu f \, \big(1 + o(\epsilon)\big).       \label{der}
\end{equation}
It is important to note and easy to see that in this approximation the
formalism has diffeomorphism invariance, that is the commutators
transform as tensors.  Indeed if we make the change of variables $\,
x^{\prime\mu}= x^{\prime\mu}(x^\rho)$, using (\ref{der}) we obtain
\begin{equation}
 \kb J^{\prime\mu\nu} =[x^{\prime\mu},x^{\prime\nu}] =
\frac{\p x^{\prime\mu}}{\p x^{\rho}}
\frac{\p x^{\prime\nu}}{\p x^{\sigma}}\, [x^\rho,x^\sigma]=\kb\,
\frac{\p x^{\prime\mu}}{\p x^{\rho}}
\frac{\p x^{\prime\nu}}{\p x^{\sigma}} \, J^{\rho\sigma} .
\end{equation}
This means that the diffeomorphism  symmetry is present not only in 
the classical limit but also in the linear order in $\kbar$.

To define and study a noncommutative space $\c{A}$ there are two
equivalent paths (the 2-fold way).  Starting with a set of vector
fields $e_\alpha$ on a smooth manifold which satisfy (\ref{RicciRot})
one can define the momenta as the solutions to the equations
\begin{align}
[p_\alpha, x^\mu] &= e^\mu_\alpha     ,         \label{ema}
\end{align}
which follow from the expression
\begin{equation}
 dx^\mu =e^\mu_\alpha \theta^\alpha.
\end{equation}
Relation~(\ref{qr}) then leads to a set of important consistency
conditions which one has to solve.  Alternatively, one can start from
a moving frame $\theta^\alpha$ on a classical manifold satisfying
\begin{equation}
d\theta^\gamma = -\frac 12 C^\gamma{}_{\alpha\beta} \theta^\alpha \theta^\beta ,
\end{equation}
and search  for an analogous noncommutative frame, extending thus
$\,\Omega^*(\c{C})$ to $\,\Omega^*(\c{A})$.  In both cases
one obtains the essential ingredients to construct a noncommutative geometry, 
an algebra and a compatible differential calculus.  We find however that
there are restrictions on the ingredients which arise from the Jacobi
identities.

\subsection{Dynamics}

The commutator~(\ref{commutator}) must satisfy Jacobi
identities which can be  written in the form
\begin{align}
A_{J\rho} = \epsilon_{\rho\lambda\mu\nu}[x^\lambda,J^{\mu\nu}] = 0
                                               \label{jacobi}
\end{align}
of vanishing of an anomaly. We shall accept the assumption that this
condition is the only obstruction to the associativity of the product.
There are also Leibniz rules, obtained by replacing one position
generator $x^\lambda$ by a momentum $p_\alpha$. As we are restricting
our attention to the case when momenta depend on coordinates, the
additional Jacobi identities follow from~(\ref{jacobi}). The Leibniz
rules can be written in the form
\begin{equation}
d\big([x^\mu, x^\nu] - \kb J^{\mu\nu}\big) = 0.                            \label{D}
\end{equation}
It is easily seen that the analogous
compatibility conditions for equation~(\ref{commutator}),
\begin{equation}
d\big([x^\mu, \theta^\alpha]\big) = 0         ,                           \label{E}
\end{equation}
are equivalent in fact to the quadratic relation~(\ref{qr})
for the momenta.

\initiate
\section{Real-space foliation}

The simplest spherically symmetric
noncommutative space is the fuzzy
sphere~\cite{Hop82,Mad92a}. Anticipating that isotropic
spaces in  higher dimensions contain it as  a subspace, we  will
review its basic properties.  Two angular variables of the polar
coordinate system on the fuzzy sphere  are described by three Cartesian
coordinates which satisfy commutation relations of the algebra
of rotations.  The Lie algebra  $so_3$ has irreducible
representations by three $n\times n$ complex matrices $J^a\in M_n$
which we normalize such that
\begin{equation}
[J^a, J^b] = i \epsilon^{abc} J_c,\qquad J^2 = \tfrac 14 (n^2-1).    %   \label{abc}
\end{equation}
Let  $\,  x^a$, $a=1,2,3\, $,  be the quantized coordinates.
Quantization  is defined by relation
\begin{equation}
 x^a =\frac\kbar r J^a             ,                     \label{bohr}
\end{equation}
which is an analog of the Bohr quantization and postulates that
the area of the quantum sphere contains an integer number of elementary 
cells of area $\kbar$. It defines also the radius $r$
since from the Casimir relation of $so_3$  for large $n$ we obtain
\begin{equation}
 r^2=g_{ab}x^ax^b = \frac{\kbar n}{2}.                           \label{Casimir}
\end{equation}
The commutator of the coordinates is then given by
\begin{equation}
 [x^a,x^b] = \kb C^{ab}{}_c x^c =\frac{\kb}{r}\, \epsilon^{abc} x_c.      \label{fs}
\end{equation}
Relation (\ref{Casimir}) suggests that the three-dimensional  
space in polar coordinates is, or can be represented as a direct sum
\begin{align}
{\cal A}_3 = \bigoplus_n M_n  
\end{align} 
of all irreducible representations of $so_3$: this model
is called the `onion model'.
%  But the onion model
%  cannot be  nontrivially  (with nonsingular 
% $J^{\mu\nu}$) extended to four dimensions because of 
% the no-go theorems on  the extensions of the simple Lie algebras. 
% We are here not restricted only to the noncommutative spaces which 
% have the structure of  Lie algebra; nonetheless the Jacobi identities 
% impose tight constraints. A further problem is that, as $r$  is
% identified with the Casimir operator, there is no room for 
% the corresponding  momentum $\, p_r$. 
As we need a four-dimensional manifold we shall not use it;
 simple extensions by one algebra generator
do not have the desired properties and 
limits too~\cite{Buric:2008th,Buric:2009zz}.
Therefore we add the radius and the time 
as  coordinates independent on $x^a$.
 That is, we consider  space-time 
$\,\c{A}=\c{A}^\prime\otimes\c{A}^\pprime\,$
to be the tensor  product of the fuzzy-sphere algebra
$\c{A}^\prime$ and an algebra $\c{A}^\pprime$ generated by $r$ and
$t$.

The momenta on the fuzzy sphere  can be identified with the coordinates,
\begin{equation}
\kb p_a = \delta_{ab} x^b.
\end{equation}
We have therefore
\begin{equation}
[p_a, p_b] = C_{ab}{}^{c} p_c , 
\qquad C_{abc} = \frac 1r \,\epsilon_{abc}      \label{Pabc}
\end{equation}
and if we let $\, 
 \pi^a_b = \delta^a_b - \dfrac{1}{r^2}\,x^ax_b \, $
be the radial projection onto the sphere, we obtain from  definition~(\ref{df})
 the expression for the differentials $dx^a$,
\begin{equation}
 dx^a = \frac 1r\, \epsilon^a{}_{bc} x^b\theta^c ,
\end{equation}
and inversely,
\begin{equation}
\pi^a_b \theta^b =- \frac 1r\,\epsilon^{a}{}_{bc}x^b dx^c 
+\frac{i\kbar}{r^2}\, \delta^a_c dx^c .
\end{equation} 
In the leading order in $\kbar$ we can write
\begin{equation}
 dx^a = C^a{}_{bc} x^b\pi^c_d \, \theta^d, 
\qquad
\pi^a_b \theta^b =- C^{a}{}_{bc}x^b dx^c.
\end{equation}

\subsection{The momentum algebra}

Since the information about symmetries is contained in the
frame derivatives $e_\alpha$ that is in $p_\alpha$, the momentum 
algebra\footnote{We refer to the algebras
  generated by the coordinates and the momenta respectively as the
  position algebra and the momentum algebra although we assume
in this section  they are one and the same algebra.}
is the best starting point to search for $\c{A}$. It has in addition
a fixed quadratic form which reduces the number of possible
Ans\"atze.  We investigate then the extensions of the $so_3$ 
algebra by two operators $p_0$ and $p_4$
which we introduce in place of $r$ and $t$ as generators of $\c{A}^\pprime\, $.
We start with an algebra of a general form
\begin{eqnarray}
 && [p_a,p_b] =\frac{ i\epsilon}{2}\, \epsilon_{ab}{}^c\,
( \Pi p_c +p_c \Pi)\label{a11}\\[4pt]
 && [p_0,p_c] =\frac{ i\epsilon}{2} \,( \pi_0 p_c +p_c \pi_0)\\[4pt]
 && [p_4,p_c] =\frac{ i\epsilon}{2}\, ( \pi_4 p_c +p_c \pi_4)\\[4pt]
 && [p_0,p_4] ={ i\epsilon}\,  \Xi\\[4pt]
 && [\Xi, p_c] =\frac{ i\epsilon}{2}\, ( \Upsilon p_c + p_c \Upsilon)      \label{a12}
\end{eqnarray}
which manifestly respects rotational symmetry.
In accordance with the requirement that the algebra be quadratic
we write
\begin{eqnarray}
 && \Pi = a+bp_0 +cp_4        \\[4pt]
&&\pi_0 = a_0 + b_0 p_0 +c_0 p_4     \\[4pt]
&& \pi_4 = a_4 + b_4 p_0 + c_4 p_4 .
\end{eqnarray}
The $a_k$, $b_k$ and $c_k$ are constants while $\,\Xi $\, 
is a general quadratic polynomial in  $p_0$ and $p_4$.
When we impose  Jacobi identities on (\ref{a11}-\ref{a12})
we obtain various restrictions. For example we find
\begin{equation}
 \Upsilon  =(b_0+c_4)\,\Xi, \qquad   c\,\Xi = \frac 12( \pi_0 \Pi+\Pi\pi_0) ,
 \qquad b \,\Xi =-\frac 12 (\pi_4 \Pi +\Pi\pi_4 ) ,
\end{equation}
and from these relations we conclude that $\pi_0$ and $\pi_4$ are
mutually proportional,
\begin{equation}
 b\pi_0 +c\pi_4 =0  .
\end{equation}
The constants  are also not  independent: we have
\begin{equation}
 ba_0 + ca_4=0,\qquad bb_0 +cb_4=0,\qquad bc_0 + cc_4 =0.
\end{equation}
The most important implication of the  Jacobi
constraints is that there exists always a linear combination of  
$ p_0$ and $p_4$ which commutes with $p_a$: 
it is in fact equal to $\Pi$. Using $\Pi$,
the momentum algebra simplifies to
\begin{equation}
 \begin{array}{ll}
 ×[p_a,p_b] = i\epsilon \epsilon_{ab}{}^{c}\, \Pi p_c,       
&[\Pi,p_c] =0                 \\[6pt]
 [\pi_4,p_c] = \dfrac{i\epsilon}{2}\, c (\pi_4p_c +p_c \pi_4) , \qquad
&  [\pi_4,\Pi] = \dfrac{i\epsilon}{2}\, c (\pi_4\Pi+\Pi \pi_4)    
\end{array}
     \label{4}
\end{equation}
where we substituted
\begin{equation}
\frac{bc_0}{c} - b_0  \to c  .
\end{equation}
We adopt here the convention that the momenta are antihermitian
operators. 

It is easy to rewrite algebra (\ref{4}) in the tensor-product
form by introducing the hermitian frame components $\zeta^a$ of a
vector (which we will later use as coordinate),
\begin{equation}
 \zeta_a = \Pi^{-1} p_a           .                 \label{zeta}
\end{equation}
We have
\begin{equation}
 \begin{array}{ll}
   [\zeta_a,\zeta_b] = \kb \epsilon_{ab}{}^{c}  \zeta_c, \qquad \qquad
&[\Pi,\zeta_a] =0 \\[6pt]
 [\pi_4,\zeta_a] = 0,  &  [\Pi,\pi_4] = \dfrac{\kb c}{2} (\Pi \pi_4+\pi_4 \Pi) .  
 \end{array}                     \label{two}
\end{equation}
We see that  $\c{A}$ indeed contains an
$so_3$ as a subalgebra;  $\c{A}$ is a tensor product of  the
$so_3$ generated by three $\zeta_a$ and a quadratic
$\c{A}^\pprime$  defined by relation
\begin{equation}
\Pi \pi_4 = q\pi_4 \Pi ,                               \label{qquad}
\end{equation}
with
\begin{equation} 
 q=\frac{2+\ep c}{2-\ep c}.                        \label{q}
\end{equation}
However, differential calculus defined by $(p_a,\Pi,\pi_4)$ 
is not a tensor-product calculus because the momenta do not 
belong to the factor algebras.  This property is desired as 
otherwise for the metric we would obtain a simple product-metric.

We mentioned already that within the frame formalism  the
momentum algebra determines  space-time geometry.
 Of course there  is a freedom in the choice of the 
linear conection, which then uniquely 
gives the torsion and the curvature; these expressions one can show are
generically quadratic in the momenta, \cite{Mad00c,Buric:2009ss}.  
But momenta are {\it a priori} 
unknown functions of coordinates. Therefore although our model is 
essentially fixed, to understand it in more details and to find its 
classical limit we need to determine coordinates.

\subsection{The position algebra}

The functional dependence of $p_\alpha$ on $x^\mu$ is not unique, but
as everything else it is constrained by the algebraic structure of
$\c{A}$ and the required properties of the commutative limit. In
the present case there is only one spatial 3-vector in the algebra
so it is natural to assume that  coordinates $x^a$
are proportional to it.
In fact  we already introduced the hermitian  $\zeta^a$, 
$\, p_a = \Pi \zeta_a$, as generators
which provide the tensor-product form of the algebra (\ref{two}).   
We can thus choose $\zeta^a$ as coordinates, $x^a =\zeta^a$;
one can see easily from (\ref{yyy}) that the choice of the 
proportionality factor does not influence the form of the angular 
part of the line element. Of course, only two of the
$\zeta^a$ are independent because $ \zeta^2=\zeta^a\zeta_a$ 
is the Casimir of ${\cal A}$.
The operators $\Pi$ and
$\pi_4$  have to be mutually independent functions of
$r$ and $t$. An almost obvious Ansatz for these functions is
\begin{equation}
i\epsilon \Pi= F(r) = \frac 1r \, , \qquad  i\epsilon \pi_4 = G(t).    \label{ass}
\end{equation}

Coordinate components of the (inverse) metric 
can be found from relation 
\begin{equation}
 g^{\mu\nu} =e^\mu_\alpha e^\nu_\beta \eta^{\alpha\beta}  ,
\end{equation}
where the frame components $\, e^\mu_\alpha$ are given by
$\, e^\mu_\alpha(x)=[p_\alpha, x^\mu].$
One can also use the inverse $\, \theta^\alpha_\mu $ of $\, e^\mu_\alpha $,
\begin{equation}
 g_{\mu\nu} =\theta_\mu^\alpha \theta_\nu^\beta \eta_{\alpha\beta} .  \label{metric}
\end{equation}
Two given expressions are equivalent within the
precision we are working with, that is to the
operator-ordering ambiguities. We can also write the line element as
\begin{equation}
 ds^2 = \eta_{\alpha\beta}\, \theta^\alpha\otimes \theta^\beta ,
\end{equation}
and then express the frame 1-forms through the differentials,
$ \,   \theta^\alpha = \theta^\alpha_\mu (x)\,  dx^\mu  $,
to obtain $g_{\mu\nu}$.
The constant frame metric $\, g_{\alpha\beta}=\eta_{\alpha\beta}$
here has the signature \, $(-++++)\,$. 

The commutator of $r$ and $t$,
\begin{equation}
[r,t]=\kb J(t,r)             \label{tr}
\end{equation}
can be obtained from compatibility of  (\ref{4}) with 
(\ref{tr}). Using it we find easily
\begin{equation}
J =- c\mu^2 r\,\frac{ G}{\dot G}.                    \label{assump}
\end{equation}
For  differentials of the coordinates we have
\begin{eqnarray}
 && d\zeta^i = [p_b ,\zeta^i]\theta^b
 +[\Pi,\zeta^i]\theta^0 +[\pi_4,\zeta^i]\theta^4
= \frac 1r\, \delta^{ia} \epsilon_{abc} \zeta^b \theta^c  \\[4pt]
&& dr =  cr G\, \theta^4             \\[4pt]
&& dt =  \frac{c}{r}\frac{ G}{\dot G}\, 
(\delta_{ia} \zeta^i\theta^a +\theta^0),
\end{eqnarray}
we introduced  $\, \zeta^i =\delta^i_a\zeta^a$.
From these expressions we obtain the  frame components
\begin{equation}
 e^i_c =\frac 1r\, \delta^{ia} \epsilon_{abc} \zeta^b,\qquad e^4_4 =  c r G,
\qquad e^0_a =\frac{c}{r}\frac{ G}{\dot G}\, \zeta_a \qquad 
e^0_0 =  \frac{c}{r}\frac{ G}{\dot G}  .
\end{equation}
We can calculate the metric by applying  (\ref{metric});
the nonzero elements are
\begin{eqnarray}
&& g^{ij} =e^i_a e^j_b \delta^{ab} 
=\frac {1}{r^2}(\delta^{ij}\delta^{bd}-\delta^{id}\delta^{jb})\zeta_b\zeta_d
= \frac{1}{r^2}\,\delta^{i}_a\delta^j_b \zeta^2\pi^{ab}               \label{gmn}
\\[4pt] &&
 g^{44} = \left( c r G \right)^2
\\[4pt] &&
  g^{00} =   - (1- \zeta^2)\left( \frac{c}{r}\frac{ G}{\dot G} \right)^2
\\[2pt] &&
 g^{0i} = \kb \, \frac{c}{r^2}\frac{ G}{\dot G}\, \zeta^i  .
\end{eqnarray}

We see immediately that in the classical limit $\kbar\to 0 $ 
off-diagonal terms $\, g^{0i}$ vanish.  Further, since we can
assume that the Casimir $\zeta^2 =\frac 14 \,\epsilon^2 (n^2-1)$ is 
small, we have  $1-\zeta^2>0$. The angular part of 
the metric is, as expected, proportional to the projector on the 
sphere  $\pi^{ab} $,
\begin{equation}
 \pi^{ab} = \delta^{ab}-\frac{\zeta^a\zeta^b}{\zeta^2}.
\end{equation}
 The prefactor in (\ref{gmn}) is also  correct:
the inverse of $g^{ij}$ is on the sphere given by   
\begin{equation}
 g_{ij} = \frac{r^2} {\zeta^2}\, \delta^a_i\delta^b_j\,  \pi_{ab},     \label{yyy}
\end{equation}  
and therefore the angular part of the line element is
\begin{equation}
{ds}_{\Omega}^2= g_{ab}\, d\zeta^a d\zeta^b  =
r^2 d\Omega .                                     \label{ang}
\end{equation}

In the classical limit we obtain
\begin{equation} 
 ds^2 = -\frac{r^2}{c^2(1-\zeta^2)} \frac{\dot G^2}{G^2} \,dt^2 +
\frac{1}{c^2 r^2 G^2}\, dr^2 + r^2 d\Omega.
\end{equation}
We have a time-dependent metric.  Namely, we cannot assume that
$G(t)=G_0$, a constant, because momenta $\Pi$ and $\pi_4$ have to
be functionally independent.  Introducing instead of $t$ a new variable $\tau$
\begin{equation}
 \frac{\dot G}{G}\, dt =- \lambda d\tau,\qquad G=e^{-\lambda\tau},
\end{equation}
we can simplify the line element to 
\begin{equation}
 ds^2 = -\frac{\lambda^2 r^2}{c^2(1-\zeta^2)}  \,d\tau^2 +
\frac{1}{c^2 r^2 } e^{2\lambda\tau}\, dr^2 + r^2 d\Omega,    \label{cosmo}
\end{equation}
 with the curvature scalar $R$
\begin{equation}
  R=\frac{2(1+c^2 -c^2\zeta^2) }{r^2} -12 c^2 e^{-2\lambda\tau}.
\end{equation}
This metric, though spherically symmetric is obviously  not
homogeneous. The universe expands differently in different directions:
if we define the Hubble parameter $H$ as the rate of expansion in the
radial direction we find that $H=\lambda$. The $\lambda$ 
on the other hand is related to the noncommutativity of space-time, 
\begin{equation}
[ r,\tau] = - \frac{ \kb c}{\lambda} \, r  .
\end{equation}
An interesting change of variables,
 $\eta = G(t)$, $\, \chi = \dfrac 1r \,$, transforms the line
element (\ref{cosmo}) to an almost conformally-flat form,
\begin{equation}
 ds^2 =\frac{1}{c^2 \eta^2\chi^2}
\left( -\frac{1}{1-\zeta^2}\, d\eta^2 +d\chi^2 +c^2\eta^2 d\Omega\right) .
\end{equation}

Since the Ansatz (\ref{ass}) which we used does not
give a static metric which could be interpreted as
 an extension of the Schwarzschild black hole, 
perhaps  it is possible to modify it to obtain the extension of
the FRW cosmology?  Then instead of (\ref{ang}) we would need
 the angular line element  $ \,
ds^2_\Omega = a^2(t) r^2 d\Omega \, $. This corresponds presumably
to a more general dependence,
\begin{equation}
 i\epsilon\Pi =F(r) N(t),\qquad i\epsilon\pi_4 = L(r) G(t).  \label{Ans}
\end{equation}
Let us analyse this possibility leaving 
the commutator Ansatz the same, (\ref{tr}).  A comparison of (\ref{Ans})
with  (\ref{4}) gives the compatibility equation
\begin{equation}
- \frac{\dot N}{N}\,\frac{L^\prime}{L} +
 \frac{\dot G}{G}\,\frac{F^\prime}{F}  = \frac{c\mu^2}{I_0I_4} .     \label{*3}
\end{equation}
As variables $r$ and $t$ are  separated it is not difficult 
to find a solution to the last equation. We have
\begin{eqnarray}
-\frac{F^\prime}{F}=\frac{\alpha}{I_4},\qquad
-\frac{L^\prime}{L}=\frac{\beta}{I_4},\qquad
 \frac{\dot N}{N}=\frac{\gamma}{I_0},\qquad
 \frac{\dot G}{G}=\frac{\lambda}{I_0},                           \label{eqs}
\end{eqnarray}
where  constants $\, \alpha$, $\beta$, $\gamma$ and $\lambda$ 
satisfy
\begin{equation}
 \beta\gamma-\alpha\lambda=c\mu^2\neq 0.                     \label{cons}
\end{equation}
The change the variables to
\begin{equation}
 \rho(r) ={- \int \frac{dr}{I_4(r)}}, \qquad
 \tau(t) ={\int \frac{dt}{I_0(t)}},
\end{equation}
drastically simplifies all equations and we can solve (\ref{*3}):
\begin{eqnarray}
 F=e^{\alpha\rho},\qquad L=e^{\beta \rho},  \qquad
 N=e^{\gamma\tau},\qquad G=e^{\lambda\tau}.                   \label{fgLG}
\end{eqnarray}
% The solution to (\ref{*3}) can be  slightly  generalised:
% for example,  instead of (\ref{eqs}) we can use the solutions
% to equations
% \begin{eqnarray}
% \gamma \frac{L^\prime}{L}- \lambda
%  \frac{F^\prime}{F}=\frac{c\mu^2 }{I_4},\qquad
%  \frac{\dot N}{N}=\frac{\gamma}{I_0},\qquad
%  \frac{\dot G}{G}=\frac{\lambda}{I_0}  .
% \end{eqnarray}
% This generalisation is however not too significant.

Now it is possible to calculate the metric.  From the differentials
\begin{eqnarray}
 && d\zeta^a =FN \epsilon_{abc}\zeta^b\theta^c   \\[4pt]
&&dr = - J\left( F\dot N\zeta_b\theta^b +F\dot N \theta^0 +L\dot G\theta^4    \right)    \\[4pt]
&&dt = J \left( F^\prime N\zeta_b\theta^b +F^\prime N \theta^0 
+L^\prime G\theta^4    \right)  ,
\end{eqnarray}
we obtain  the nonvanishing components of the frame,
\begin{equation}
\begin{array}{lll}
 e^i_c =FN \delta^{ia} \epsilon_{abc}\zeta^b \quad & & \\[4pt]
 e^4_a = - JF\dot N\, \zeta_a, & e^4_0 =- JF\dot N ,\quad &  e^4_4 =- J L\dot G   \\[4pt]
 e^0_a = JF^\prime N\, \zeta_a,& e^0_0 = JF^\prime N , & e^0_4 = J L^\prime G .
\end{array}
\end{equation}
Using (\ref{metric}) we  find:
\begin{eqnarray}
 && g^{ij} =(FN)^2 \zeta^2 \delta^i_a\delta^j_b \, \pi^{ab}  \\[4pt]
&&g^{44} =J^2\left( -(F\dot N)^2 (1-\zeta^2) +(L\dot G)^2 \right) \\[4pt]
&&g^{00} =J^2\left( -(F^\prime N)^2 (1-\zeta^2) +(L^\prime G)^2 \right) \\[4pt]
&&g^{04} = J^2\left( F^\prime N F\dot N (1-\zeta^2) - L^\prime G L\dot G \right)  \\[4pt]
&&g^{0i} = - \kb JFF^\prime N^2 \zeta^i        \\[6pt]
&&g^{4i} = \kb JF^2 N\dot N  \,\zeta^i .
\end{eqnarray}
Again the off-diagonal components $g^{0i}$ and $g^{4i}$ vanish in the
commutative limit and the angular part is proportional to the
projector on the sphere: the corresponding classical metric is
spherically symmetric, non-static and block-diagonal.  Inverting the
angular part we get
\begin{equation}
g_{ab} = (FN)^{-2}\zeta^{-2}\, \pi_{ab} ,
\end{equation}
that is,
\begin{equation}
ds_\Omega^2 = g_{ab} \, d\zeta^a d\zeta^b = (FN)^{-2} d\Omega .
\end{equation}
The $(r,t)$ part on the other hand gives
\begin{eqnarray}
 &&g_{00} =\frac{1}{c^2 I_0^2}\left(
\frac{\gamma^2 }{(LG)^2}-\frac{\lambda^2}{ (1-\zeta^2)(FN)^2}  \right) \\[4pt]
&&g_{44} =\frac{1}{c^2 I_4^2}\left(
\frac{\alpha^2 }{(LG)^2}-\frac{\beta^2}{ (1-\zeta^2)(FN)^2}  \right) \\[4pt]
&&g_{04} = \frac{1}{c^2I_0 I_4}\left(
\frac{\alpha\gamma }{(LG)^2}-\frac{\beta\lambda}{ (1-\zeta^2)(FN)^2}  \right) .\label{04}
\end{eqnarray}
One can  further simplify  using 
 variables $\rho$ and $\tau$ and obtain
\begin{equation}
ds_{\rho}^2 =
 - \,\frac{1}{ c^2(1-\zeta^2)(FN)^2}\, (\lambda d\tau +\beta d\rho)^2 +
\frac{1}{c^2(LG)^2}\, (\gamma d\tau +\alpha d\rho)^2 .             \label{one}
\end{equation}
Having the line element written as this
it is easy to recognize that the change of coordinates
\begin{equation}
 FN \to \frac 1r, \qquad LG \to e^{-\lambda \tau}
\end{equation}
gives back the  metric (\ref{cosmo}). This is a manifestation of 
diffeomorphism invariance of the formalism, and shows 
that the noncommutative space which we obtained is in fact unique.

\subsection{A representation}

The coordinates of the solution~(\ref{ass})  satisfy the
 commutation relation
\begin{equation}
 e^{\lambda \tau} r = q^{-1}\, r e^{\lambda\tau}.       \label{last}
\end{equation}
Parameter $ q $ defined in~(\ref{q}) is unitary because the
constant $ c$ is real, while  $r$ and $\tau$ are hermitian
operators. Equation (\ref{last}) can be rewritten as
\begin{equation}
  e^{\lambda \tau} r e^{- \lambda\tau} = r +\lambda [\tau,r] + 
\frac{\lambda^2}{2!} [\tau,[\tau,r]] + \dots =q^{-1}r  ,
\end{equation}
and we see that it holds for
\begin{equation}
 [\tau,r] =ikr                                        \label{Iar}
\end{equation}
with $\, \tan ({ k\lambda}/{2}) ={\epsilon c}/{2}$:
the algebra  is formally equivalent to the Heisenberg algebra.
%: to see this we start
% with the latter, given by $x$ and $y$ with $\, [x,y]= i \, $ and we
% introduce 
% \begin{equation}
% \tau= \frac k2\, (xy - \frac{i }{2} ), \qquad r = y^2  ,                \label{f}
% \end{equation}
% which satisfy (\ref{Iar}).  
The spectrum of $\tau$ is
the real line and that of $r$ the positive real line.
A representation on the Hilbert space of square integrable 
functions of one variable $s$ is  given by
\begin{equation}
\tau =  \frac {ik}{4}\,( s \, \frac {d}{ds} +  \frac {d}{ds} \, s),
\qquad  r= s^2 ,
\end{equation}
where $\tau$ is the dilatator, or by
\begin{equation}
 \tau = ik\,\frac{d}{ds},\qquad r =e^s.
\end{equation}
We have here an indication of the importance of the calculi
in the description of the geometries: the differential calculus 
usually introduced on a space described by 
the  Heisenberg algebra is  flat, while here  the subspace 
$(r,\tau)$ has a constant negative curvature, \cite{BurMad05b}.  
One can find even an example of an algebra over which there are two 
different calculi with geometries having as commutative limit two different
topologies.

\initiate
\section{Internal-space foliation}

We have seen that 
in the previous section geometry of the space-time was fully
determined by momenta and by their commutators.  
Such situation is  typical when we apply the noncommutative 
frame formalism in its `minimal' version, that is 
when all $p_\alpha\in \c{A}$.  A close relation between the
algebraic and the geometric structures is however a
general feature of any  noncommutative geometry, also in cases when
we need to extend the space-time in order to obtain 
the appropriate phase space or the prescribed symmetries.

We shall attempt now to find static spherically symmetric metric 
starting from the algebra of coordinates. This means that we will 
first make an Ansatz for the position commutators and for the frame,
and then analyse the implied consistency relations. This approach is perhaps 
more intuitive and seems easier because the position algebra is not
restricted in its form like the momentum algebra. However,
the overall number of equations remains  the same if we constrain
the momenta to belong to $\c{A}$: a real advance comes 
if we allow for some of the derivations to be external.
As a result, we shall find a way to extend $\c{A}$ and to
obtain static solutions.

\subsection{The position algebra}

Assume that $\c{A}$ is generated by operators $\, x^\mu=(\xi^a,\rho,r,t)$
with for some large  $n$
\begin{align}
\xi^a = \dfrac{2}{n}  J^a.
\end{align}
Variable $\rho$ is a fifth generator which we can consider as a Kaluza-Klein
extension. We expect a general spherically symmetric solution to
depend on  $\rho$, $r$ and $t$ but we shall  restrict our attention
to the static case.

The multiplication table is
\begin{alignat}{2}
&[\xi^a,\xi^b] = \dfrac{2i}{n}\, \epsilon^{abc}\, \xi_c, \qquad\quad
&&[\xi^a,\rho]= [\xi^a,r]=[\xi^a,t]=0             \label{1a}
\\[6pt]
&[\rho,t] = \kb J^0 \rho, %   \qquad
&&[r,t] = \kb J                         \label{0a}\\[6pt]
&[\rho,r] = \kb J^4 \rho          \qquad \quad
&&                 \label{rhot}
\end{alignat}
and to insure spherical symmetry we assume that
$ J=J(\rho,r,t)$, 
$J^0=J^0(\rho,r,t)$, $ J^4=J^4(\rho,r,t)$. 
The position algebra (\ref{1a}-\ref{0a}) is restricted by Jacobi
identities. The nontrivial one is
\begin{equation}
 [\rho,[r,t]] + [r,[t,\rho]] + [t,[\rho,r]] = 0,
\end{equation}
and it gives to first order the equation
\begin{equation}
 J^0 \dot J - J\dot J^0 
+  J^4 J^\prime - J J^{4\prime} 
+ J^4\rho \p_\rho J^0 - J^0\rho \p_\rho J^4 = 0,
\end{equation}
or if we introduce $\gamma_0$ and $\gamma_4$ as
$\, J^0 = J\gamma_0,$ $ \, J^4 = J\gamma_4 $,
\begin{equation}
 \dot \gamma_0  +\gamma_4^{\prime} 
+ \gamma_0\rho \, \p_\rho \gamma_4 - \gamma_4\rho\, \p_\rho \gamma_0 = 0.
\end{equation}
As a convenient solution we can choose
\begin{equation}
 J^4=0  ,   \quad \gamma_4=0               ,                \label{j4}
\end{equation}
which  implies 
\begin{equation}
\dot\gamma_0 = 0      .                                    \label{**}
\end{equation}
The last equation is identically fulfilled in the static case.  

For comparison with the previous section we introduce
$x^a=\rho\xi^a$, $x^a x^b \delta_{ab}=\rho^2$ and the intermediate 
variable $L = {2\rho^2}/{n\kbar} \, $. Then we have
\begin{equation}
 [x^a,x^b] = \dfrac{i\kbar}{\rho}\, L\epsilon^{ab}{}_{c} \, x^c .
\end{equation}

\subsection{The  frame}

As the angular part of the frame we choose the Mauer-Cartan frame 
of the group multiplied by a function $h$ to account for the change 
in volume of the 3-sphere as we move along the radial; the radial
 and the time-like components of the frame are, we suppose, diagonal.
We obtain for the frame in five dimensions
\begin{equation}
\begin{array}{ll}
 \theta^a = -{h}{\rho}^{-1}\, \epsilon^{a}{}_{bc}\, x^b dx^c 
+\rho^{-2} \,x_b \theta^b x^a ,
       \qquad & 
dx^a =({h\rho})^{-1}\, \epsilon^{a}{}_{bc}\, x^b \theta^c   \\[6pt]
\theta^4 = gdr    , &    dr = g^{-1} \theta^4                         \\[6pt]
 \theta^0 = f dt  ,  &          dt = f^{-1} \theta^0        .  
\end{array}      \label{d1} 
\end{equation}
The $f$, $g$ and $h$ are functions of $\rho$, $r$ and $t$.
As on the fuzzy sphere this Ansatz gives 
\begin{equation}
 \rho\, d\rho + d\rho \, \rho =0
\end{equation}
which is unusual because it implies
 $\, d\rho =0 \,$  in linear order even though $\rho$ is not in 
the center. In particular, $d\rho =0\, $ 
in the classical limit.  This is a specific  feature of our calculus
(\ref{d1}) and is a consequence of the fact that the algebra
of coordinates, we shall see, is different from the algebra
of momenta. 

To be consistent,  differential calculus (\ref{d1}) has to be
compatible with algebra (\ref{1a}-\ref{0a}) and this compatibility is
expressed by equations (\ref{D}).  To leading order the right-hand
side of these equations is given by
\begin{equation}
 d J^{\mu\nu} = \p_\rho J^{\mu\nu} dx^\rho =
\p_\rho J^{\mu\nu} \, e^\rho_\alpha \theta^\alpha,
\end{equation}
while to calculate the left-hand side we  use  the
property that the elements of the frame basis commute with  ${\cal A}$,
\begin{equation}
 [dx^\mu, x^\nu] =[e^\mu_\alpha \theta^\alpha, x^\nu]
= [e^\mu_\alpha, x^\nu]  \theta^\alpha .
\end{equation}
An analysis of all  constraints which follow from these
requirement yields:
\begin{equation}   \label{Comp}
 \begin{array}{lll}
 \dot L =0    , \quad &
 L^\prime =0 , &\ \ 
 h^\prime J^4 + \dot h J^0 =0   \\[6pt]
\dot J^4=0 , \quad  &
J^{4\prime} + {g}^{-1}  (g^\prime J^4 +\dot g J^0) =0,
 &\ \  (h +\rho \p_\rho h)  J^4
- \dot h J =0   \\[6pt]
J^{0 \prime} =0,  \quad
& J^{0\prime} +{f}^{-1} (f^\prime J^4 
+\dot  f J^0)=0,
& \ \ ( h+\rho \p_\rho h)  J^0
+h ^\prime J =0  
\\[6pt]
 &\dot J + f^{-1} ( \dot f J 
-\rho \p_\rho J^4)=0 ,
\quad & \ \  J^\prime + g^{-1} ( g^\prime J 
+ \rho \p_\rho J^0 ) =0 .
 \end{array}
\end{equation}
We must impose additional relations which assure that the exterior
multiplication of two frame 1-forms is well defined. These
relations follow from the definition of the wedge product~\cite{Mad00c}
\begin{equation}
 \theta^\alpha \theta^\beta = P^{\alpha\beta}{}_{\gamma\delta}
\theta^\gamma \theta^\delta        .    \label{*}
\end{equation}
In lowest order we obtain
\begin{equation}
 [\theta^{\alpha}, \theta^{\beta}] = \kb\mu^2 Q^{\alpha\beta}{}_{\gamma\delta}
\theta^\gamma \theta^\delta       ,                          \label{FE}
\end{equation}
where $\, Q^{\alpha\beta}{}_{\gamma\delta}\, $ are constants introduced
in (\ref{pq}). In our case, (\ref{FE}) become the following equations:
\begin{eqnarray}
&&\hskip-2cm [\theta^a,\theta^b]=0                     \label{20}\\[6pt]
 && \hskip-2cm [\theta^0,\theta^0] =i\kbar \,fg^{-1}
{\p_r}\Big(f^{-1} f^{\prime} J+
\rho f^{-1}  \p_\rho f J^0\Big)\theta^4\theta^0         \\[6pt]
 && \hskip-2cm [\theta^4,\theta^4] = -i\kbar \, gf^{-1}
{\p_t}\Big( g^{-1} \dot g\, J-
\rho g^{-1}  {\p_\rho}g\, J^4\Big) \theta^0\theta^4           \\[6pt]
&& \hskip-2cm [\theta^0,\theta^4]+(fg)^{-1} [f,g]\, \theta^4\theta^0
 = i\kbar\,
\p_t\Big( g^{-1} g^{\prime} J+\rho
 g^{-1}  \p_\rho g\,J^0\Big)\theta^0\theta^4                \\[6pt]
&&\hskip-2cm \phantom{[\theta^0,\theta^4]+(fg)\p_\rho [f,g]\, 
\theta^4\theta^0}
= - i\kbar\,
{\p_r}\Big( f^{-1} \dot f J -
\rho f^{-1} \p_\rho f  J^4\Big)\theta^4\theta^0   .     \label{24}
\end{eqnarray}
There is one  remaining relation for $\, [\theta^a,\theta^0]\,$ which can be
obtained from 
\begin{eqnarray*}
&& [dx^a,dt]  = \frac{1}{\rho  hf}\,\epsilon^{a}{}_{bc} x^b\, [\theta^c,\theta^0] 
+\frac{1}{\rho h^2 f^2}\,\epsilon^{a}{}_{bc} x^b \,[h,f]\theta^c\theta^0 \\[6pt]
 &&=- \frac{i\kbar}{\rho  h f}\,\epsilon^{a}{}_{bc} x^b\,Jf^{-1}
(\gamma_4 f^\prime
+\gamma_0 \dot f)\theta^c\theta^0 -  \frac{i\kbar}{fg}\, x^a\,
{\p_r}\Big( Jf^{-1} (\gamma_4 f^\prime +
\gamma_0 \dot f)  \Big)\theta^4\theta^0 
\\[6pt]  
&& =  \frac{i\kbar}{\rho hf}\,\epsilon^{a}{}_{bc} x^b\,{\p_t}
\Big( Jh^{-1}
(h^\prime +\gamma_0\rho {\p_\rho}h)\Big)
 \theta^0\theta^c
+ \frac{i\kbar}{\rho hg}\,\epsilon^{a}{}_{bc} x^b\,
{\p_r}\Big(  Jh^{-1}
(h^\prime +\gamma_0\rho {\p_\rho}h)\Big) \theta^4\theta^c .
\end{eqnarray*}

The first set of equations (\ref{20}-\ref{24}) gives the following constraints:
\begin{eqnarray}
 &&  fg^{-1}
{\p_r}\Big(  Jf^{-1}( f^\prime +
 \gamma_0 \rho \,{\p_\rho}f)\Big) = C\\[8pt]
 &&  gf^{-1}
{\p_t}\Big(  Jg^{-1}( \dot g-
 \gamma_4 \rho\,{\p_\rho}g )\Big) =C_2  \\[8pt]
&&({fg})^{-1}[f,g]
+  \kb {\p_t}\Big(Jg^{-1} (g^\prime+
\gamma_0\rho\, {\p_\rho}g)\Big)  
  =({fg})^{-1}\,[f,g]
+ \kb {\p_r}\Big( Jf^{-1}( \dot f -
\gamma_4\rho \, {\p \rho}f)\Big) =C_3,\nonumber
\end{eqnarray}
and from the second set we obtain:
\begin{eqnarray}
 &&\hskip-6cm  fg^{-1}
{\p_r}\Big(  Jh^{-1}( h^\prime +
 \gamma_0 \rho \, {\p \rho}h)\Big) =C_4  \label{C4} \\[8pt]
 && \hskip-6cm 
{\p_r}\Big(  Jf^{-1}( \gamma_4 f^\prime
 +\gamma_0 \dot f)
\Big) =0   \\[8pt]
&& \hskip-6cm  ({hf})^{-1}[h,f]-
  \kb \Big(  Jf^{-1} (\gamma_4 f^\prime+
\gamma_0 \dot f)\Big)     =C_5 , 
\end{eqnarray}
the $C_k$ are constants.

\subsection{Solutions}

We obtained a relatively complicated set of equations,
but as we are looking for static solutions we can assume that
no function depends on time. In fact, the proposed choice
 $\, J^4 =0$,  $\, L= {2\rho^2}/{n\kbar} \, $
satisfies almost all of the equations. The first nontrivial
constraint is $\, J^{0\prime}=0\, $ and it implies
 $\, J^0=J^0(\rho) $. The remaining are:
\begin{eqnarray}
&&\gamma_0 h+ h^\prime +\gamma_0\rho {\p_\rho}h =0      \label{23}
\\[6pt]
 && J^{-1} J^\prime + g^{-1} ( g^\prime
+\gamma_0 \rho {\p_\rho}g ) =0 \\[6pt]
&& fg^{-1}
\Big(  Jf^{-1}( f^\prime +
 \gamma_0 \rho \, {\p_\rho}f)\Big)^\prime = C ,            \label{25}
\end{eqnarray}
whereas  (\ref{C4}) 
is  satisfied identically with $C_4=0$. We have 
three equations for three  frame functions $f$, $g$ and $h$.

When we are solving~(\ref{23}-\ref{25})  we should not
forget the diffeomorphism invariance. There are essentially 
two different cases. In the simplest case the frame
does not depend on $\rho$, and equations reduce to 
ordinary differential equations.
We can define the  radial coordinate arbitrarily, for example by
fixing the frame function as  $\,g=1$; or $\, h=r$;  or  $\, fg=1$.
Let us briefly review these choices and the corresponding solutions.

For $\, g=1$ the radius $r$ is the geodesic normal coordinate and  
(\ref{23}-\ref{25}) become:
\begin{equation}
 g=1,\quad J={\rm const},\quad \gamma_0={\rm const}=\gamma,
\qquad
f (\log f)^\pprime ={\rm const}, \quad \gamma +(\log h)^\prime =0.
\end{equation}
The  solution is
\begin{equation}
 h = h_0 e^{-\gamma r },
\qquad f = f_0(\cosh 2\beta r + 1) ,
\end{equation}
where $\beta $, $h_0$ and $f_0$ are the integration constants.
% For the small values of $r$ we have
% \begin{equation}
%  h  \simeq h_0(1 - \gamma r) ,
% \qquad f  \simeq 2f_0 +2 f_0\beta^2 r^2,
% \end{equation}
% with an interesting flat limit  $\, h_0  \to 0$, $\, h_0 \gamma \sim 1$,
% $ \, f_0 \sim \dfrac 12$. 
The corresponding classical limit is easily derived,
\begin{eqnarray}
ds^2 = -f^2 dt^2 + dr^2 + h^2 d\Omega  =- f_0^2 \, (\cosh 2\beta r + 1)^2 dt^2 + dr^2 +
h_0^2\,  e^{-2 \gamma r}\, d\Omega  ,           \label{111}
\end{eqnarray}
and it has the scalar curvature 
\begin{equation}
 R=-8\beta^2 -6\gamma^2+\frac{2}{h_0^2}\, e^{2\gamma r} +4\beta\, \frac{\beta +\gamma \sinh 2\beta r}{\cosh^2\beta r}.
\end{equation}
If radius $r$ is defined such that the area of the sphere be $\, 4\pi r^2$, 
that is $\, h=r$, we obtain 
\begin{equation}
 h=r,\quad   \gamma_0 =\frac{ J_0}{J} =-\frac 1r,
\qquad \frac 1r +(\log g)^\prime =0,\quad
f\big(r(\log f)^\prime\big)^\prime =-\frac{C}{J_0} \, g.
\end{equation}
Denoting the integration constant of the third equation by $\gamma$ we find
\begin{equation}
 g=\frac{1}{\gamma r},
\end{equation}
and clearly we have the same metric as (\ref{111}) expressed in
variable $ \,\bar r = h_0 e^{-\gamma r} \to r\, $.
We  further get
\begin{equation}
        f=\big(\frac{r}{h_0}+\frac{h_0}{r}\big)^2 
\end{equation}
as a solution for $f$ corresponding to nonvanishing 
$\, C=-8h_0\mu J^0\gamma\neq 0$.
For $\, C=0$  the solution is
\begin{equation}
 f=\beta r^\alpha.
\end{equation}
In particular for values  $\alpha =1$ and $\beta =\gamma\,$ the 
metric has  the Schwarzschild form, $f=g^{-1}$.
\newline
For the third definition of $r$,  $\, fg=1$, the equations become
\begin{equation}
fg=1,\quad 
(\log Jg)^\prime =0 ,\quad
f^2 f^\pprime =C, \quad
\gamma_0 +(\log h)^\prime =0 .
\end{equation}
We have
\begin{equation}
 g=f^{-1}, \quad J=\alpha f,\quad 
\log  h =-\frac{ J^0}{\alpha} \int  f^{-1} dr ,
\end{equation}
where $f$ obeys
\begin{equation}
\frac 12 \, \frac{d{f^\prime}^2}{df }= \frac{C}{f^2}.            \label{int}
\end{equation}
A solution to the last equation is for example
\begin{equation}
 f =\left(\gamma r+\beta\right)^p\qquad {\rm for}\ \ p=1, \, \frac 23\, 
\end{equation}
but the integral (\ref{int}) can be as well solved in general 
and yields the implicit solution
\begin{equation}
\gamma r +\beta =\frac{2C}{\gamma^2}\left(-\frac{x}{x^2+1} +\arctan x\right) ,
\quad  f=\dfrac{2C}{\gamma^2}\,\frac{x^2}{x^2+1} \, .
\end{equation}
We have therefore in case when the
metric does  not depend on $\rho$ three families 
of solutions: esentially they represent one noncommutative 
space, except  perhaps for singular values of the integration constants
when the spaces might differ.  This space is 
static and curved and, since everything depends on just one variable $r$,
 has a straightforward classical limit.

What happens when we include the dependence on $\rho\,$ in the frame functions? 
We have seen that $\rho$ is essentially quantum variable so  perhaps we can 
expect solutions which are physically more interesting. 
 The set of equations~(\ref{23}-\ref{25}) 
is in this case a set of  partial
equations in $r$ and $\rho$.\footnote{Note that derivatives 
$\p_\rho$ in these equations only come from evaluating commutators 
with $\rho $.} This implies that
we have many more solutions, and we shall here restrict to
 the case when  both conditions $\, h=r$ and $\ fg=1$ are fulfilled.
The equation for $h$ then gives 
\begin{equation}
 \gamma_0 =-\frac 1r,\qquad J=- J_0 r,           \label{values}
\end{equation}
where $J_0$ is an arbitrary function $J_0=J_0(\rho)$ of $\rho$.
Imposing  $\, fg=1$ we obtain that the remaining 
two equations are consistent for $\, C=0$ and 
\begin{equation}
 (r \p_r -\rho \p_\rho) \log f =1  .              \label{ddd}
\end{equation}
 We have already found a particular solution to this equation, 
$f =\gamma r$; we need  therefore in addition to solve
 the corresponding homogeneous equation
\begin{equation}
 (r \p_r -\rho \p_\rho) \log F =0 ,
\end{equation}
which is easy: the solution is an arbitrary function $\, F=F(\rho r)$.
Thus the general solution to (\ref{ddd}) is
\begin{equation}
 f=g^{-1}=\gamma r \, F(\rho r)     ,              \label{form}
\end{equation}
with the corresponding metric
\begin{equation}
 ds^2 =-f^2 dt^2 +\frac{1}{f^2} \, dr^2 + r^2 d\Omega.
\end{equation}
Although solution (\ref{form}) is restricted in its form, 
 in the classical limit it gives practically all static metrics
because clasically,  $\rho$ as a constant. 
For example, taking 
\begin{equation}
 F(x) = \sqrt{ \frac {1}{x^2} -\frac{2m\gamma}{x^3}  }
\end{equation}
and assuming that  $\rho=\gamma$  we obtain the Schwarzschild metric.

\subsection{A  representation}

Noncommutative space $\c{A}\, $  is a tensor product 
$\c{A}=\c{A}^\prime\otimes\c{A}^\pprime$ of a first
factor generated by the $\xi^a$ and a second factor generated by
$\rho$, $r$ and $t$.  To represent it we
choose a Hilbert space $\c{H}$ which is a tensor product
$\c{H}=\c{H}^\prime\otimes\c{H}^\pprime$, with $\c{H}^\prime$ 
given by a representation $J^a$ of $so_3$ and
$\c{H}^\pprime = L^2(\b{R}^2)$.  To be explicit we consider 
the case defined by (\ref{values}) and $J_0=1$; the  commutation 
relations  are
\begin{equation}
 [\rho,t] = i\kbar \rho,   \qquad\ 
[r,t] = - i\kbar r            ,         \qquad \ 
[\rho,r] = 0.
\end{equation}
In this particular case the coordinates satisfy a Lie bracket relation and
the corresponding Lie algebra is a singular contraction of
$sl_2(\b{R})$: the five position generators are closely related to the
six generators of the Lie algebra of the Lorentz group.
The $\c{A}^\pprime$ can be represented for example on the Hilbert space
 $\c{H}^\pprime$ of square integrable functions of two variables $(s,u)$,
in analogy with the representation discussed in the previous section, 
\begin{equation}
 \begin{array}{ll}
  \xi^a \mapsto \dfrac{2}{n} J^a \otimes 1 \otimes 1, \qquad
&  \rho  \mapsto 1\otimes  s^{-2} \otimes u
\\[8pt]
t  \mapsto   1 \otimes \dfrac{i\kbar}{4}( s\dfrac{d}{ds}+  
\dfrac{d}{ds}s) \otimes 1  ,
\qquad
& r  \mapsto 1 \otimes  s^2 \otimes 1.
 \end{array}
\end{equation}

\initiate
\section{Conclusions}

We found in this paper two families of noncommutative spherically
symmetric geometries which can be considered as extensions of 
static and cosmological solutions to the Einstein equations.
The problem of extending of a classical spherically symmetric
geometry to a noncommutative space (associative algebra $\c{A}$)
might seem at the first sight easy. 
However, the simplest way of extension, the tensor 
product, is not satisfactory because no classically relevant metric 
is a product metric. Therefore one has to extend both the algebra
and the associated differential calculus nontrivially, and that
gives nontrivial set of constraints.

Following the intuition that gravity is related to geometry
we used a noncommutative version of the Cartan frame formalism 
\cite{Mad00c}. The main input in this formalism is the set of tetrads or 
the frame. In the basis of the frame 1-forms the metric components are 
constants and the differential is defined naturally,
with respect to the given or required gravitational configuration.
However the algebraic structures are rigid and 
there are constraints: first, the Jacobi constraints in the algebra 
and second, the consistency constraints necessary for the
compatibility  of the algebra with the calculus.
 One of the significant features of the formalism is
the possibility to generalize the diffeomorphism invariance.
To first order in noncommutativity  this invariance is exactly the 
same as in the Einstein gravity and one is allowed to
choose the most convenient set of coordinates.

Our concrete problem  was
 formulated as: how to extend the fuzzy sphere 
to a four-dimensional static or cosmological 
space? We first started by extending the symmetry, 
that is the momentum algebra, assuming at the same time
that the frame derivations are inner. This assumption gave 
an essentially unique solution for $\c{A}$: an algebra 
generated by five coordinates $\, (\zeta^a,r,t)$ 
or by five momenta $\, (p_a,\Pi,\pi_4)$ constrained by one relation
that is, possessing one Casimir operator, $\zeta^2$. This algebra with
the associated calculus describes a spherically symmetric
nonstatic space, which however is not spatially homogeneous
that is, not isotropic. Nonetheless the space has some interesting
properties for example the Hubble parameter is given
by noncommutativity. Everything  in the model was 
essentially  fixed: the Jacobi constraints 
behaved as field equations.

In the second part of the paper we extended the 
original algebra by adding one variable,
 so $\c{A}$   was generated by six coordinates 
$(\xi^a,\rho,r,t)$  constrained by one constraint, 
or by five unconstrained coordinates $(x^a,r,t)$. 
In this setup, as we have seen, one could obtain
practically all static spherically symmetric 
configurations as  classical limit. The price to pay 
was an additional variable $\rho$ with somewhat unusual 
properties: the differential calcus implied that
 $d\rho =0$ in the classical limit while 
$\rho $ was not a  constant (we have for example $[\rho,t]\neq 0$).
In principle one would consider the existence 
of an element of the algebra which is not a constant  and
nevertheless has vanishing differential an undesirable feature.
However  such elements
are not uncommon, we mention as example the dilatator $\Lambda$ of 
the quantum line. The exact role 
of $\rho$ in this specific case remains to be understood better,
along with the question of the appropriate representation; 
in any case the classical  relation $d\rho=0$ suggests that 
a natural way to interpret $\rho$ is as a Kaluza-Klein parameter 
which measures the internal space.
In addition, one can  rather easily see that in this model
the momenta cannot belong to  $\c{A}$. Namely, 
solving equations (\ref{ema}) for $p_\alpha$ we
obtain the solution $\, i\kbar p_a = \delta_{ab}\xi^b$, 
$\, p_0 =p_0(\rho,r)$, but the remaining 
$p_4 =p_4(\rho, r, t)$ is  inconsistent with the assumption 
$\, e^a_4=0$ which gives diagonal and static metric.

Momentum operators are,  in classical gravity, always external, 
$\, p_\alpha = e^\mu_\alpha \p_\mu$. In addition
when we deal  with commutative space, the Jacobi identities and the 
de Rham consistency conditions are trivially satisfied.
In the opposite, maximally constrained noncommutative case
when the phase and the configuration spaces are identical,
the Jacobi identities and the compatibility  conditions 
practically fix the dynamics and the geometry. Here we observe that,
 with increase of the phase space,  constraints become less
restrictive and we have more freedom to choose solutions:
more noncommutative geometries can be defined consistently.
It is certainly possible to continue along these lines and
find other interesting quantum spaces with nondegenerate 
noncommutativity, correct symmetries and 
desired commutative limits, and that is what we plan to 
investigate in our future work.

\vskip0.5cm
\begin{large}
{\bf Acknowledgement}
\end{large}
This work was supported by the Serbian Ministry of Education, Science
and Technological Development Grant ON171031. The authors worked on
the subject especially while visiting on several occasions ESI in
Vienna and AEI in Berlin. They would like to take this opportunity to
thank Harald~Grosse, Hermann~Nicolai and Stephan~Theisen for their 
hospitality.

% 
% 
% \newpage
% 
% \bibliographystyle{utphys}
% \bibliography{abbrev,refgen,refmad,proceedings}

\end{document}